\documentstyle[epsfig,amsmath,amssymb,graphicx,tabularx]{elsart}
\newcounter{saveeqn}

\def\gsimeq{\,\,\raise0.14em\hbox{$>$}\kern-0.76em\lower0.28em\hbox  
{$\sim$}\,\,}  
\def\lsimeq{\,\,\raise0.14em\hbox{$<$}\kern-0.76em\lower0.28em\hbox  
{$\sim$}\,\,}  
\def\beqy{\begin{eqnarray}}
\def\eeqy{\end{eqnarray}}
\def\bmlet{\begin{mathletters}}
\def\emlet{\end{mathletters}}
\begin{document}
 \begin{frontmatter}  
 
\title{Large-scale QRPA calculation of $E1$-strength and its impact on the neutron
capture cross section}
\author{S. Goriely$^1$, E. Khan$^2$}
\address{
$^1$Institut d'Astronomie et d'Astrophysique, ULB - CP226, 1050 Brussels, Belgium \\
$^2$ Institut de Physique Nucl\'eaire, IN$_{2}$P$_{3}$-CNRS, 91406 Orsay,
France}
%
\begin{abstract}
Large-scale QRPA calculations of the $E1$-strength are performed as a first
attempt to microscopically derive the radiative neutron capture cross
sections for the whole nuclear chart. A folding procedure is applied to the
QRPA strength distribution to take the damping of the collective motion into
account. It is shown that the resulting $E1$-strength function based on the
SLy4 Skyrme force is in close agreement with photoabsorption data as well as
the available experimental $E1$ strength at low energies. The increase of
the $E1$-strength at low energies for neutron-rich nuclei is qualitatively
analyzed and shown to affect the corresponding radiative neutron capture
cross section significantly. A complete set of $E1$-strength function is
made available for practical applications in a table format for all $8 \le
Z\le 110$ nuclei lying between the proton and the neutron drip lines.
\end{abstract}
\end{frontmatter}
\section{Introduction}
The radiative neutron captures by exotic nuclei are known to be of
fundamental importance in the rapid neutron-capture process (or
r-process) invoked to explain the origin  of approximately one half of
the nuclides heavier than iron observed in nature. The r-process is
believed to take place in environments characterized by high neutron
densities ($N_n \gsimeq 10^{20}~{\rm cm^{-3}}$), so that successive
neutron captures can proceed into neutron-rich regions well off the
$\beta$-stability valley. If the temperatures or the neutron densities characterizing
the r-process are low enough to break the $(n,\gamma)-(\gamma,n)$ equilibrium, the
r-abundance distribution depends directly on the neutron capture rates by the
so-produced exotic neutron-rich nuclei \cite{go98}. The neutron capture rates
are commonly evaluated within the framework  of the statistical model of Hauser-Feshbach.
This model makes the fundamental assumption that the capture process takes place with the
intermediary formation of a compound nucleus in thermodynamic equilibrium. In this
approach, the Maxwellian-averaged $(n,\gamma)$ rate at temperatures of relevance in
r-process environments strongly depends on the electromagnetic interaction, i.e the
photon de-excitation probability. 

The photon transmission coefficient from a compound-nucleus excited
state is dominated by the $E1$ transition
which is classically estimated within the Lorentzian representation of
the giant dipole resonance (GDR), at least for medium- and heavy-mass
nuclei. Reaction theory relates the
$\gamma$-transmission coefficient for excited states to the ground state
photoabsorption assuming the giant resonance to be built on each excited
state. Experimental photoabsorption data confirms the simple
semi-classical prediction of a Lorentzian shape at energies around the
resonance energy $E_{GDR}$. For this reason, for practical applications involving
neutron-capture cross section calculations, the $E1$-strength function is
exclusively estimated so far on the basis of a form or another of the Lorentzian
model. To explain the measured strength functions and $\gamma$-ray
intensities \cite{mc81}, it is however required to bring correction to the low-energy
behavior of the strength. This is achieved by introducing an energy-dependent
width \cite{mc81} or an energy- and temperature-dependent width
\cite{go98,ko87,mu00} guided by more detailed finite Fermi system descriptions
\cite{ka83} of the $E1$-strength function at energies below the neutron
separation energy $S_n$. These corrections were shown to
improve significantly the predictions of the experimental radiation widths
and gamma-ray spectra \cite{go98,ko90}. 

  In addition to the generalized Lorentzian model \cite{go98,mu00,ko90}, improved
description  of the $E1$-strength function can be derived from the
semi-classical thermodynamic pole approach  \cite{pl02}, the microscopic random
phase approximation (RPA) calculations (based on Hartree-Fock \cite{ca97} or
relativistic mean field
\cite{vr01} ground state descriptions) or the theory of finite Fermi systems
\cite{ka83}. None of the so-called microscopic approaches provides at the moment a
complete set of $E1$-strength function for practical applications, so that global
predictions for unknown nuclei can only be taken from phenomenological approximation of
Lorentzian type.  However, extrapolating these approximations (including the
location of the resonance energy) to exotic neutron-rich nuclei might not be reliable.
In particular, Catara et al.
\cite{ca97}
 showed in the framework of RPA calculations that the neutron
excess affects the spreading of the isovector dipole strength, as well as the centroid
of the strength function. The energy shift is found larger than the one accounted by the
usual
$A^{-1/6}$ or $A^{-1/3}$ dependence given by the phenomenological liquid drop
approximations. 
Various calculations \cite{ca97,vr01,va92,su90} on the exotic neutron-rich nuclei
predict  that the existence of a neutron mantle could introduce a new type of collective
mode corresponding to an out-of-phase vibration of the neutron-proton core
against the neutron mantle. The restoring force for this soft dipole
vibration is predicted to be smaller than that of the GDR.  The total $E1$ strength in
this so-called pygmy resonance (PR) is small (around a few percent of the total GDR
strength), but, if located well below the neutron separation energy,  can significantly
increase the radiative neutron capture cross section \cite{go98}. Van
Isacker et al. \cite{va92} estimated the PR energy and strength relative
to that of the GDR  within the incompressible fluid model of
Goldhaber-Teller, assuming the  dynamics of the  dipole oscillations to
be governed by the neutron-proton interaction. Such calculations were then
systematically applied to all the neutron-rich nuclei and shown to
affect significantly (up to factors of about 100) the neutron capture rates of the $2
\lsimeq S_n[{\rm MeV}] \lsimeq 4$ closed shell nuclei \cite{go98}. Large uncertainties
in the description of the PR (in particular its energy and strength) remain and only
more fundamental descriptions could shed light on its existence, as well as its
relative importance and impact on neutron capture rates. 
In addition, different measurements suggest that some particular
enhancement of the $E1$-strength could be located at low energies even on stable nuclei,
a feature that cannot be predicted by the phenomenological approach. In particular,
dipole transitions to bound states investigated by means of the nuclear resonance
fluorescence confirmed the systematic existence of a pygmy $E1$-resonance at energies
below the neutron separation energy \cite{gov98}. PR are
observed in $fp$-shell nuclei as well as in heavy spherical nuclei near closed shells
(Zr, Mo, Ba, Ce, Sn and Pb). 

For various type of applications, and more particularly for nucleosynthesis
applications, but also  accelerator driven systems, a great effort must be made to
improve the reliability of the $E1$-strength predictions, in particular for
experimentally unreachable nuclei and in the low-energy region (typically below the
neutron separation energy).  Generally speaking, the more microscopic the underlying
theory, the greater will be one's confidence in the extrapolations out towards the
neutron-drip line, provided, of course, the predictions are capable of reproducing
experimentally known data with a sufficient degree of accuracy. For this reason, use
should be made preferentially of microscopic or semi-microscopic global predictions based
on sound and reliable nuclear models which, in turn, can compete with more
phenomenological highly-parameterized models in the reproduction of
experimental data. Microscopic approaches (like the RPA) are almost never
used for practical applications, because of the numerical difficulty
associated with large-scale predictions and the fine tuning of the model to reproduce
accurately experimental data, especially when considering a large data set. We
present here the first attempt to construct a complete set of $E1$-strength
function derived from Quasi-particle RPA (QRPA) calculations based on an effective
nucleon interaction of Skyrme-type. In Sect.~2, the QRPA calculations are presented.
Different Skyrme interactions are compared in their predictions of the $E1$-strength and
GDR energies. In Sect.~3, the QRPA shortcoming in broadening the $E1$-strength is cured
by folding the QRPA strength on a Lorentzian-type function. A folding
prescription is provided to reproduce experimental photoabsorption and available
experimental $E1$ strengths at low energies, in particular average resonance capture
(ARC) data. Section~4 is devoted to the analysis of exotic neutron-rich nuclei and the
impact of the newly-determined strength on the neutron-capture cross sections and rates
of interest for astrophysical applications. Conclusions are given in Sect.~5

\section{QRPA calculations of the $E1$-strength}
         Microscopic models based on the RPA are well known for their
	 accuracy to describe collective modes such as giant resonances
	 \cite{ng83}. Due to the mean field approximation, they are able to
	 calculate the strength function on the whole nuclear mass range in
	 a short amount of time. An improved RPA model is the QRPA which
	 takes the pairing effect into account. This effect is expected to
	 be important for open-shell nuclei. In our case the ground state is
	 calculated with a Hartree-Fock+BCS (HF+BCS) model. It should be
	 noted that it is the first time that large scale calculations on
	 more than 6000 nuclei are performed in order to calculate not only
	 their ground state (with HF+BCS) but also their excited states
	 (with the QRPA). This allows to characterize the structure model on
	 the whole nuclear chart, with the effective nucleon-nucleon
	 interaction as only input, and opens the exciting perspective to
	 study the properties of the effective interaction with respect to
	 the nuclear excitations. However the BCS approximation is known to
	 not properly describe the pairing correlation for many drip-line
	 nuclei \cite{do96}. Ideally, for very neutron-rich nuclei along the
	 drip-line, the model should take into account the coupling between
	 the nuclear mean-field, the pairing and the particle continuum. Different approaches,
such as the coordinate space HFB theory \cite{en99} or the
	 continuum QRPA equations derived from the Time-Dependent-HFB theory
	 \cite{ma01,gr01} have been developed. However, large-scale calculations with such
sophisticate continuum-QRPA models are far from being feasible at the present time.
For this reason, we will restrict ourselves to the QRPA based on HF+BCS to get a
qualitative insight of this approach on the neutron capture cross
	 section by exotic neutron-rich nuclei. This study also provides interesting information
on how calculated cross-sections obtained with two different approaches compare,
namely the phenomenological Lorentzian (exclusively used nowadays in cross section
calculation) and the microscopically rooted QRPA.

        The QRPA model here employed \cite{kh00a} has already been
	successful in describing low-lying collective states in the oxygen
	\cite{kh00b}, sulfur and argon \cite {kh01} isotopic chains. The
	main feature of this model is its self-consistency in the sense that
	the mean field and the residual interaction are calculated from the
	same Skyrme effective interaction. The ground state is calculated
	within the HF+BCS approximation. The HF and the BCS equations are
	both solved in a self-consistent procedure. The only inputs to the
	calculations are the Skyrme interaction and the value of the
	constant pairing gap $\Delta$. For the sake of simplicity, we choose
	a gap given by \cite{bo69}:
\begin{equation}\label{eq:de}
        \Delta = 12~ A^{-\frac {1}{2}} {\rm MeV}~.
\end{equation}
         Because of the constant gap approximation we use a cutoff beyond
	 which subshells do not participate to the pairing effect. This
	 cutoff was chosen to take into account all the subshells of the
	 major shell where the Fermi level belongs to. The pairing gap
	 (\ref{eq:de}) is also used for magic nuclei. In this case the
	 effect is not important because of the large shell spacing, so that
	 the giant dipole resonance strength is not affected. Note that the present 
  calculation with constant pairing gap is self-consistent only in the particle-hole
  channel, but not in the particle-particle channel. On top of the HF+BCS ground
  state, the excited
	 $J^{\pi}$=1$^-$ states are calculated within the QRPA model. The
	 QRPA equations are solved in the configuration space so as to
	 exhaust the energy weighted sum rule (EWSR). All QRPA calculations
	 are performed in the spherical approximation. The deformation
	 effects are introduced in a phenomenological way as explained in
	 Sect.~3. The results on strongly deformed nuclei should therefore
	 be taken with circumspection. In E1-strength calculations a
	 well-known spurious state appears, which corresponds to a
	 translation of the whole nucleus. In our case, since we do not use
	 a projection method or a modified operator, the spurious state is
	 unambiguously identified in each E1-strength calculation : it lies
	 close to 0 MeV (in a range between 0 and 2 MeV) and takes almost
	 all the isoscalar strength (around 90\%). Once identified this
	 state is eliminated from the E1-strength.
        
\begin{figure}
\centerline{\epsfig{figure=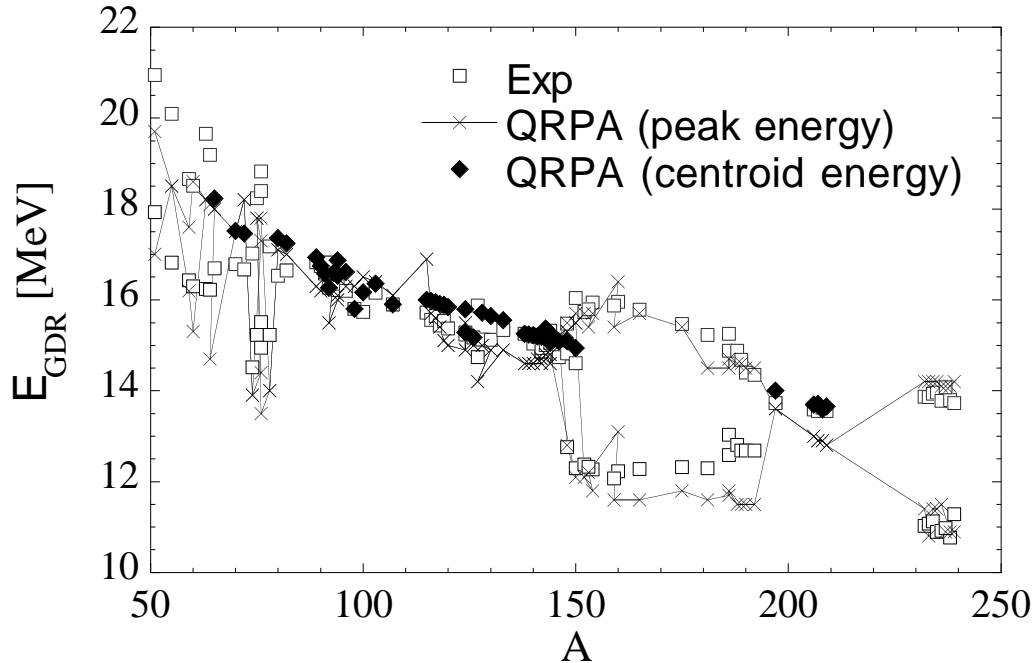,height=9.0cm,width=14cm}}
\caption{\label{fig01}{Comparison between the experimental GDR energies
\cite{di89} and the QRPA calculations obtained with the SLy4 Skyrme
force. The QRPA centroid energies (diamonds) are displayed for 48 spherical nuclei only.
The experimental GDR energies for spherical and deformed nuclei are also compared with
the GDR peak energies (crosses) obtained after folding the QRPA distribution and
including the deformation effects.}}
\end{figure}

        The most significant observable to check the validity of the model
and the accuracy of the interaction is the GDR energy. Several Skyrme forces
have been used, namely SIII \cite{be74}, SGII \cite{ng81}, SLy4 \cite{ch98}
and MSk7 \cite{go01b}. In all cases, the pairing interaction is treated
within the simple above-described prescription (Eq. (\ref{eq:de})). In order to select the
most adequate interaction to the
$E1$ calculations, we compare the calculated centroid and measured GDR
energies for 48 spherical nuclei \cite{di89}. The root mean square
deviations between the calculated centroid and measured energies are,
respectively, 2267 keV for SIII, 573 keV for SGII, 457 keV for SLy4 and 564
keV for MSk7. The results are accurate for SGII, SLy4 and MSk7 interactions,
as shown in Fig.~\ref{fig01} for the specific case of SLy4. Except for the
lightest ($A\lsimeq 80$) nuclei, all GDR energies are predicted within a few
hundred keVs. On this basis we decide to use SLy4 for the large-scale
calculation of the $E1$-strength and the corresponding radiative neutron
capture. However, it should be noted that the agreement obtained with MSk7
is also good, although for some given nuclei (in particular $^{130}$Te and
$^{133}$Cs) the GDR energy is larger than the experimental value by more
than 1~MeV. This force has an effective mass of M$^*$/M=1.05 and was fitted
exclusively to the experimental masses of 1888 nuclei, with a root mean
square error of 738 keV \cite{go01b}. This result shows that an effective
interaction like MSk7 characterized by a relatively large isoscalar and
isovector effective mass (both equal to 1.05 of the nucleon mass) can
reproduce the nuclear mass as well as GDR data. A low isovector effective mass
of M$^*_v$/M~$\simeq 0.70$ might therefore not be a necessary condition to
reproduce photoabsorption cross sections in contrast to \cite{kr80}.

\section{Folding procedure of the $E1$-strength}
\begin{figure}
\centerline{\epsfig{figure=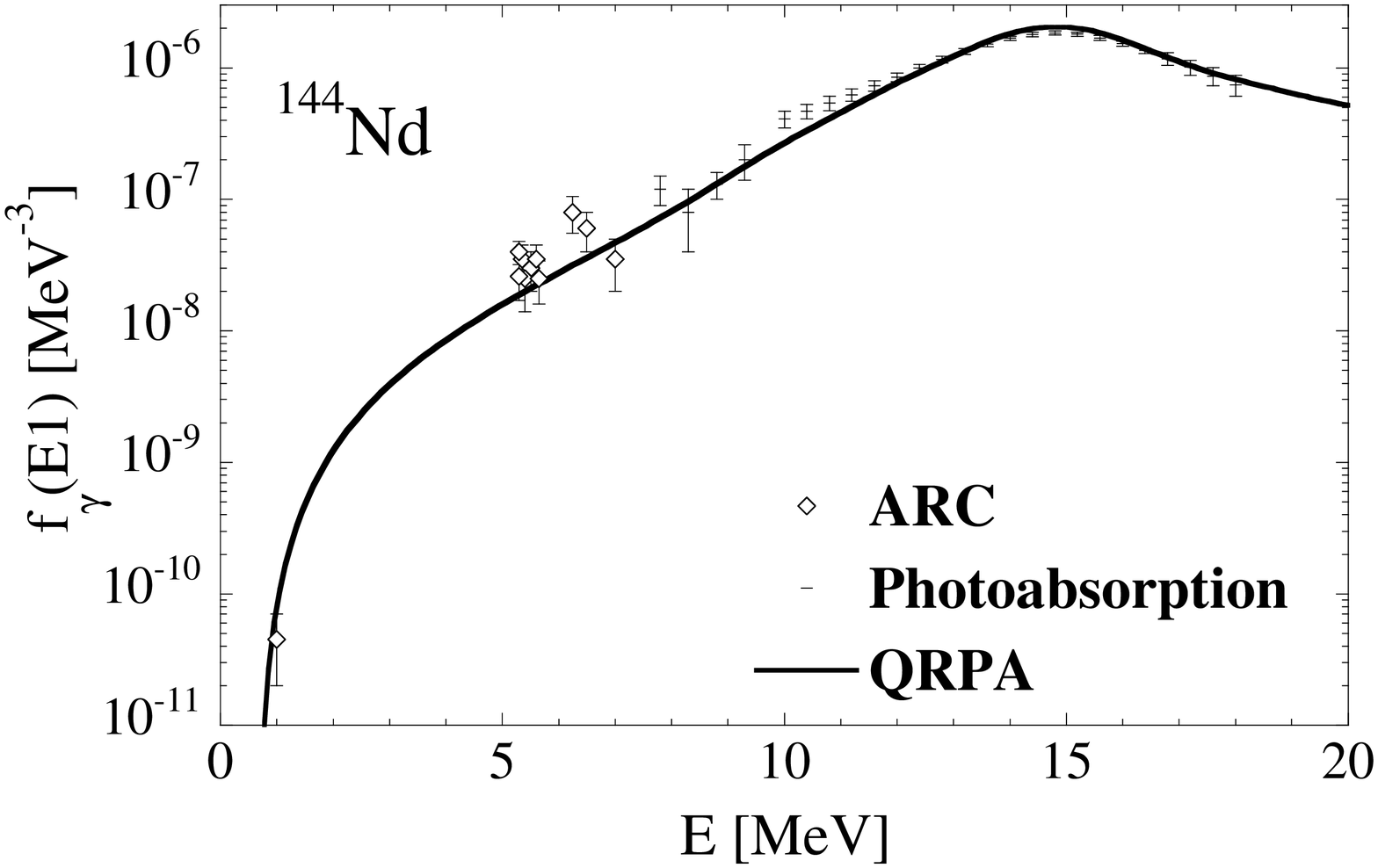,height=9.0cm,width=14cm}}
\caption{\label{fig02}{Comparison of the photoabsorption data \cite{di89} and measured
primary photon strength functions \cite{ko90} for the $^{143}$Nd(n,$\gamma$) reaction
with the QRPA predictions. For the experimental data at 1 MeV, see discussion
in \cite{ko90}. The predictions are obtained with the SLy4 Skyrme force.}}
\end{figure}

\begin{figure}
\centerline{\epsfig{figure=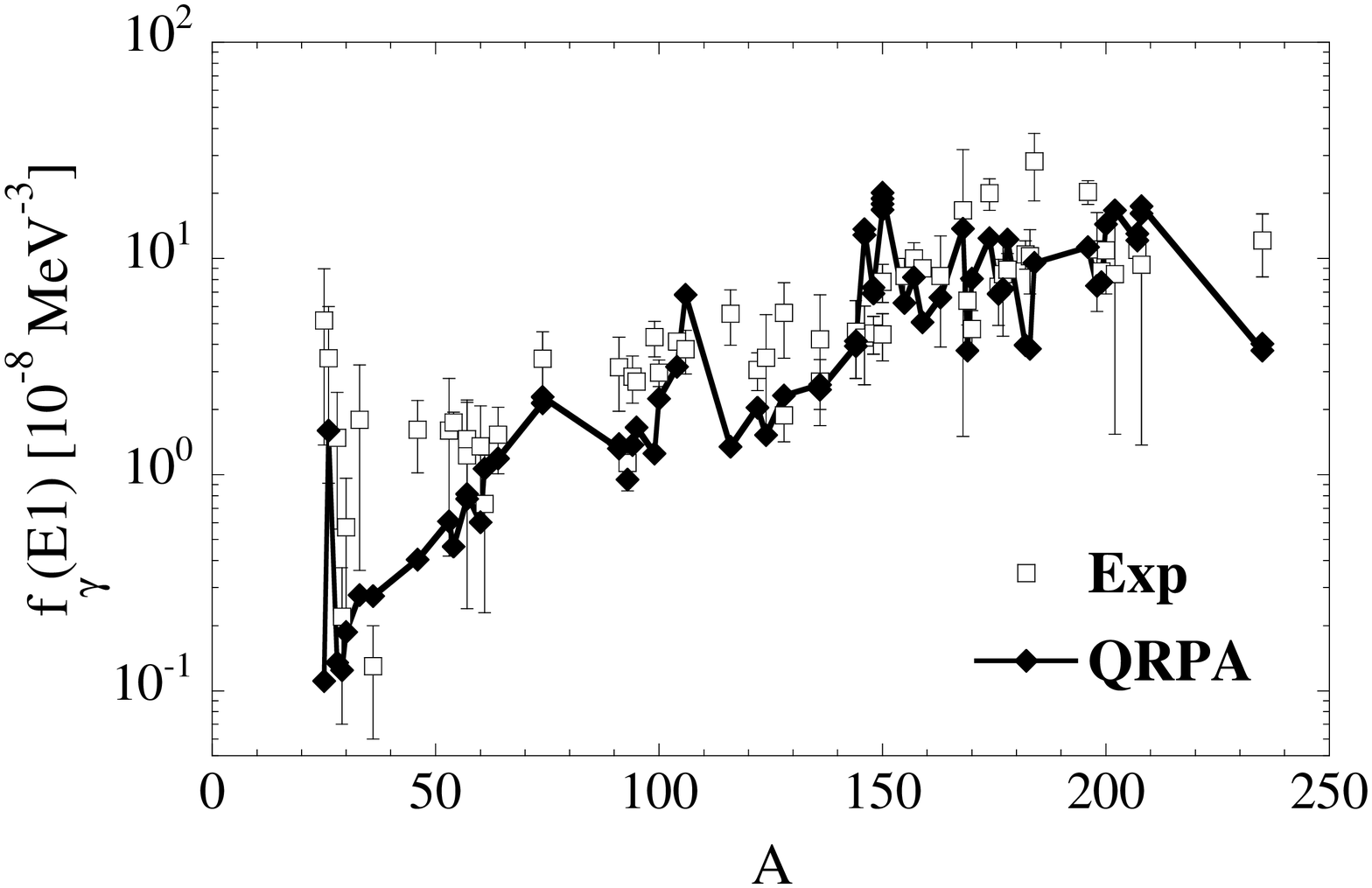,height=9.0cm,width=14cm}}
\caption{\label{fig03}{Comparison of the QRPA low-energy $E1$-strength functions
with the experimental compilation
\cite{ripl} including resolved-resonance and thermal-captures measurements, as well as
photonuclear data for nuclei from $^{25}{\rm Mg}$ up to $^{239}{\rm U}$ at energies
ranging from 4 to 8 MeV. }}
\end{figure}

As explained in Sect.~2, the QRPA provides an accurate description of the GDR centroid
and the fraction of the EWSR exhausted by the $E1$ mode. However, it
is necessary to go beyond this approximation scheme in order to describe the damping
properties of the collective motion. The GDR is known experimentally to have a large
energy width and therefore a finite lifetime. Different microscopic theories exist
(see e.g \cite{co94,co01,sc01}). However, we will restrict ourselves, for the sake of
simplicity and applicability to large-scale calculations, to an empirical broadening of
the GDR. Such a broadening is obtained by folding the QRPA strength by a normalized
Lorentzian function. Different types of Lorentzian functions as well as prescription for
the Lorentzian width are considered. The folding prescription
significantly affects the extrapolation of the $E1$-strength at low energies. In these
conditions, ARC data at low excitation energies are of particular interest to
constrain the folding procedure. The $E1$-strength is finally obtained by broadening
each QRPA resonance strength at an energy $E_i$ by the normalized Lorentzian function 
\begin{equation}
f_L(E,E_i,\gamma_i)= \frac{2}{\pi} ~\frac{\gamma_i E^2}
{(E^2-E_{i}^2)^2+\gamma_{i}^2 E^2} \quad .
\label{eq01}
\end{equation}
A  width $\gamma_i=\Gamma_{GDR}/2$ (where the GDR width $\Gamma_{GDR}$ is taken from
experimental data \cite{di89} when available or, otherwise, from empirical
systematics \cite{go98,th83}) is found to lead to results in good agreement
with experimental GDR widths and ARC data. The resulting folded GDR peak energies
obtained with the SLy4 Skyrme force are compared with experimental values in
Fig.~\ref{fig01} for spherical and deformed nuclei (see below for the treatment of
deformation). 
As shown in Fig.~\ref{fig02} in the specific
case of the spherical $^{144}$Nd nucleus, the ARC data for
the $^{143}$Nd(n,$\gamma$)$^{144}$Nd reaction \cite{ko90} are also correctly reproduced
at low energies. In addition, we compare in Fig.~\ref{fig03} the QRPA
predictions with the compilation of experimental $E1$ strength functions at low energies
ranging from 4 to 8 MeV
\cite{ripl} for nuclei from $^{25}{\rm Mg}$ up to $^{239}{\rm U}$. The data set
includes resolved-resonance measurements, thermal-captures measurements and photonuclear
data. In a certain number of cases the original experimental values need to be corrected,
typically for non-statistical effects, so that only values
recommended by \cite{ripl} are considered in Fig.~\ref{fig03}. QRPA predictions are found
to be globally in good agreement with experimental data at low energies in the whole
nuclear chart and of the same degree of accuracy as the recent calculation
within the thermodynamic pole approximation \cite{pl02}.
%

Note that the GDR width is expected to be temperature-dependent \cite{bo01}.  A
temperature-dependent  width $\gamma_i$ affects the folded
$E1$-strength function only at high temperatures ($T\gsimeq 2$~MeV) or if some
significant strength is concentrated at very low energies ($E_i
\lsimeq \gamma_i$), but does not influence the
predictions of the low-energy data shown in Figs.~\ref{fig02}-\ref{fig03}. Note also 
that introducing a temperature-dependent spreading width $\gamma_i$ in Eq.~\ref{eq01}
does not give rise to a zero $E\rightarrow 0$ limit of the $E1$-strength function due to
the specific characteristics of the Lorentzian function considered in
the folding procedure.
%

In the case of deformed spheroidal nuclei, the GDR is known to split into two major
resonances as a result of the different resonance conditions characterizing the
oscillations of protons against neutrons along the axis of rotational symmetry and an
arbitrary axis perpendicular to it. In the phenomenological approach, the Lorentzian-type
formula is generalized to a sum of two Lorentzian-type functions of energies $E^l_{GDR}$
and width $\Gamma_{GDR}^l$ \cite{th83}, such that
\begin{eqnarray}
\label{eq02}
& & E^1_{GDR}+2~E^2_{GDR}=3E_{GDR} \\ \nonumber
& & E^2_{GDR}/E^1_{GDR}=0.911 \eta + 0.089
\end{eqnarray}
where $\eta$ is the ratio of the diameter along the axis of symmetry to the diameter
along an axis perpendicular to it. In turn, when not available experimentally, the width
$\Gamma_{GDR}^l$ of each peak can be expressed as a function
of the respective energy
$E^l_{GDR}$
\cite{th83}. A similar splitting of the resonance strength is used in the folding
procedure given by Eq.~\ref{eq01},  each resonance energy
$E_i$ being divided with an equal strength according to Eq.~\ref{eq02} and characterized
by a width $\gamma_i$  proportional to $\Gamma_{GDR}^l$ following the above-mentioned
relation. Note that distributing the strength equally between the two resonance peaks 
(and not twice more on the high energy peak as done in the phenomelogical approach) is
found to give optimal location and relative strength of both
centroid energies as observed experimentally. It is shown in
Fig.~\ref{fig01} that the splitting between both energy peaks can be rather
well reproduced by the present correction. 
The strength at low energies is supposed to be
affected in a similar way, although there is no experimental indication as such. This
prescription affects the low-lying strength in a way similar to the folding procedure,
i.e to spread the strength around the corresponding energy.

\begin{figure}
\centerline{\epsfig{figure=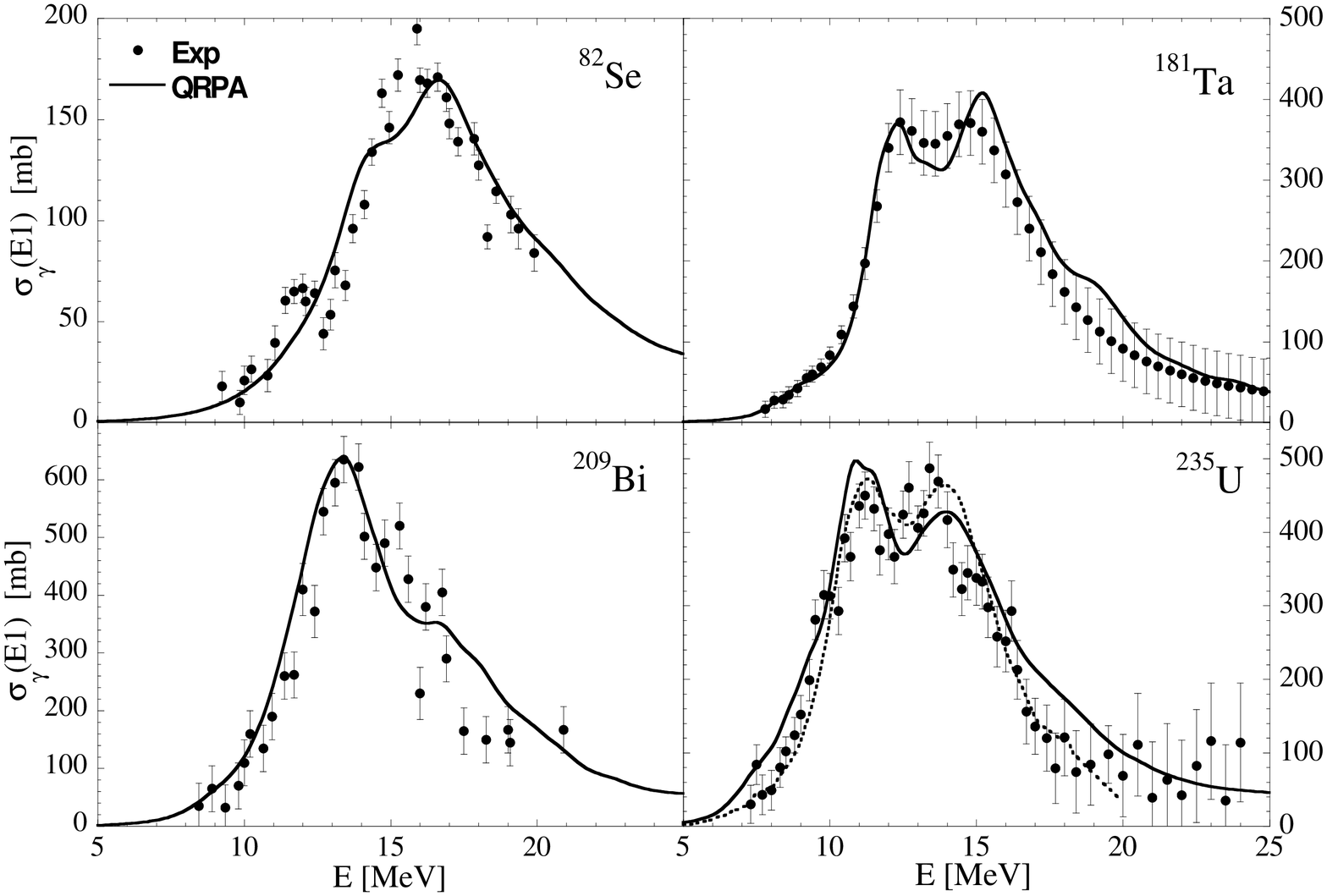,height=9.0cm,width=14cm}}
\caption{\label{fig04}{Comparison of the QRPA predictions (solid line) with the
experimental photoabsorption cross sections \cite{photo} for $^{82}{\rm
Se}$, $^{181}{\rm Ta}$, $^{209}{\rm Bi}$ and $^{235}{\rm U}$. The QRPA peak energies
were slightly renormalized according to Fig.~\ref{fig01} to reproduce the experimental
GDR energy. The dash line shown for $^{235}$U corresponds to the cross
section recommended by the Obninsk evaluated photonuclear library \cite{photo}. }}
\end{figure}

We illustrate in Fig.~\ref{fig04} how the QRPA calculation reproduces the
photoabsorption cross sections of four (spherical and deformed) nuclei in
the whole GDR region \cite{photo}. The deformation parameter $\eta$ is extracted from
the HF+BCS mass calculation of \cite{go01b}. In particular, the GDR energy and width, as
well as the double peak structure observed experimentally are rather well reproduced by
the QRPA calculation including the above-described folding and deformation prescriptions.
In the case of deformed nuclei, the relative strength between both splitted
peaks is also in good agreement with photonuclear data, in
particular for  $^{181}$Ta. Note that the double peak structure in
deformed nuclei, is not solely due to deformation effects, but also to a E1-strength
unequally distributed between 10 and 15~MeV. Although the strongest spherical QRPA
peak is located around E=13~MeV, extra strength is found in the whole $10\lsimeq E[{\rm
MeV}] \lsimeq 15$ energy range and is responsible for the asymmetries of the peaks.
Experimentally, the $E1$ distribution  is generally (at least in the
rare-earth region) characterized by a larger strength at high energies than at
low energies (see e.g.
$^{181}$Ta). However, an inversion of this GDR strength
asymmetry with a low-lying peak stronger than the high-energy one can be found in other
region and is even recommended by the Obninsk evaluated photonuclear library 
\cite{photo}, in particular for $^{234}$U,
$^{235}$U (see Fig.~\ref{fig04}),
$^{239}$Pu and
$^{241}$Pu. This recommended inversion, though not clearly observed in the 1976
experimental data of \cite{gur76} illustrated in Fig.~\ref{fig04}, is predicted by our
QRPA calculation with the above-mentionned correction for deformed nuclei. Note that
distributing unequally the strength between the two splitted peaks (for example with a
strength twice larger in the high-energy state) would not reproduce as well
experimental data as shown in Fig.~\ref{fig04}. 

 Finally, little experimental systematics exists about the exact
location and strength of the PR. Information obtained by the nuclear resonance
fluorescence technique \cite{gov98} suggests that the PR energy 
decreases with increasing masses, being around 8~MeV in the $fp$-shell region, 6.5~MeV
around $N=50, Z=50$ and 5.5~MeV for Pb. In contrast, the PR strength increases with
increasing masses, namely $B(E1)=20-60$ (in $10^{-3}$~e$^2$fm$^2$) for $fp$-shell nuclei,
$70-90$ in the $N=50$ region, $80-180$ in the $Z=50$ region, and reaching 250 for
$^{208}$Pb. Such a systematic behavior is also found by our QRPA calculation, but the
low-lying strength is  located systematically some 3 MeV higher than observed.
Improvements of the QRPA model based on the quasiparticle-phonon coupling could shift
the PR peak energy to lower values. In particular, Col\`o et al. (2001) \cite{co01} found
that the observed low-lying dipole strength in the O isotopes can be explained by the
QRPA plus phonon coupling model. Large-scale calculations in this framework are extremely
computer-time consuming and remain to be performed.

\section{Extrapolation to neutron-rich nuclei and application to the radiative neutron
capture}
\begin{figure}
\centerline{\epsfig{figure=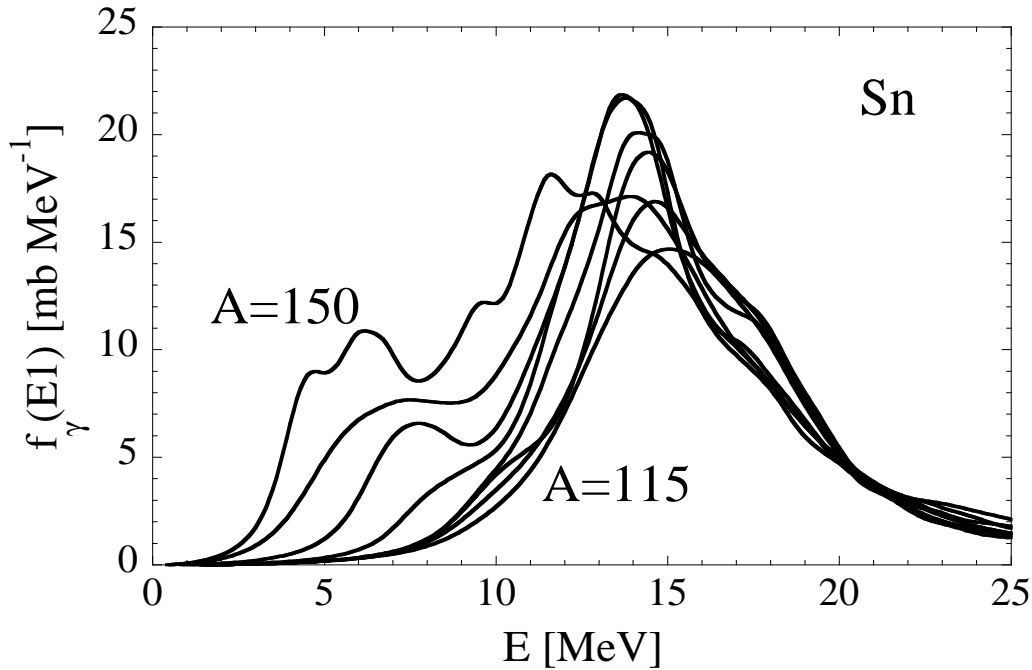,height=9.0cm,width=14cm}}
\caption{\label{fig05}{$E1$-strength function for the Sn
isotopic chain predicted by the QRPA with the SLy4 Skyrme force. Only
isotopes ranging between A=115 and A=150 by steps of $\Delta A$=5 are displayed.}}
\end{figure}

Large-scale QRPA calculations based on the SLy4 Skyrme force (See Sect.~2)
have been performed for all $8 \le Z\le 110$ nuclei lying between the proton
and the neutron driplines, i.e some 6000 nuclei. All QRPA $E1$-distributions
are folded by a Lorentzian-type function as explained in Sect.~3, although the
validity of the phenomenological folding procedure adopted in the present work cannot be
confirmed at this stage. Only further improved calculations including a detailed
description of the deformation and damping effects can shed light on this assumption.
In the neutron-deficient region, as well as along the valley of
$\beta$-stability, the resulting QRPA strength functions are very similar to
the empirical Lorentzian-like approximation and therefore in good agreement
with experimental photoabsorption data (Fig.~\ref{fig04}). When dealing with
neutron-rich nuclei, the QRPA predictions start deviating from a simple
Lorentzian shape. In particular, some extra strength is found to be located
at an energy lower than the GDR energy. The more exotic the nucleus, the
stronger this low-energy component. This is illustrated in Fig.~\ref{fig05}
for the $E1$-strength function in the Sn isotopic chain. We should stress here
 that several sound nuclear models predict different behavior for
the two-neutron separation energies of the Sn isotopes with A $>$ 132
\cite{na99}. We present here the results for our QRPA model in order to 
illustrate qualitatively the role of a microscopic description of the E1-strength for
neutron-rich nuclei. Among the 8 distributions shown in Fig.~\ref{fig05},
only the $A=150$ one corresponds to a deformed configuration responsible for
the double peak structure. For the $A\ge 140$ neutron-rich isotopes, an
important part of the strength is concentrated at low energies ($E \lsimeq
5-7$~MeV). However, it should be stressed that for nuclei close to the
drip-line, some spurious effect may be expected due to the lack of correct
continuum treatment by our model. Microscopic models such as RPA and QRPA
are well suited to interpret excitations in terms of particle-hole (p-h)
configurations, each configuration having a specific weight in the excited
state while the pairing effects allow additional configurations with holes
above the Fermi level and particles below. In this framework the low energy
component arises from two distinct origins, mainly depending on the mass of
the nucleus. For rather light or medium nuclei, one or two particular (p-h)
configurations are found responsible for the spectacular increase of the low
energy part of the $E1$-strength in very neutron-rich nuclei such as
$^{82}$Ni. For heavier elements, this increase is due to more collective
modes. For instance in the $^{258}$Pb case, around 10 (p-h) configurations
are participating to this low energy component with weights varying from 5\%
to 37\%.

 In $^{150}$Sn, mainly one (p-h) configuration is participating to the low
energy component with a 87\% weight. Phenomenological models are unable to
predict such low energy components, whatever their collectivity is. In
particular for $^{150}$Sn, all phenomenological systematics (as used for
cross section calculation) predict a $\gamma$-ray strength peaked around
15~MeV with a full width at half maximum of about 4.5~MeV
\cite{ripl} which is obviously very different from the microscopic estimate
(Fig.~\ref{fig05}). The above-described feature of the QRPA $E1$-strength
function for nuclei with a large neutron excess is found to be qualitatively
independent of the adopted effective interaction. In particular, a similar
enhanced strength is predicted by the MSk7 force which is characterized by a relatively
large nucleon effective mass. Quantitatively, the total strength at low
energies with the MSk7 interaction is smaller than in the SLy4 case.

\begin{figure}
\centerline{\epsfig{figure=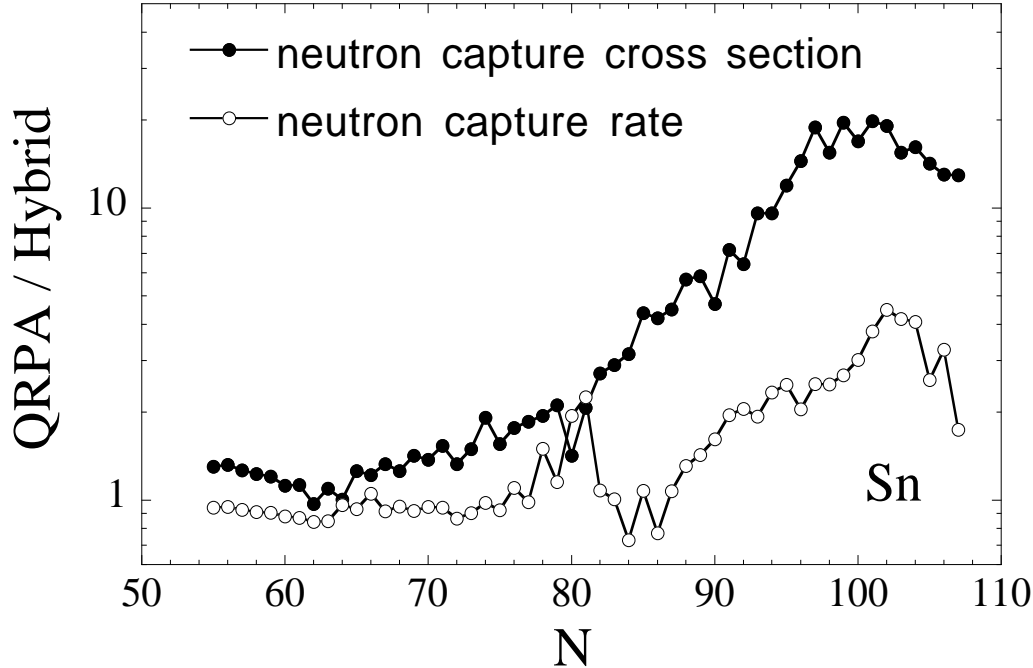,height=9.0cm,width=14cm}}
\caption{\label{fig06}{Ratios of the $(n,\gamma)$ cross section (full circles) and
Maxwellian-averaged rate (open circles) predicted with the $E1$-strength taken from the
QRPA to the one taken from the Hybrid formula \cite{go98} for the Sn isotopes. The
ratios are plotted as a function of
the neutron number. The cross section is estimated at an incident energy $E=10$~MeV
and the Maxwellian-averaged rate at a temperature of $1.5~10^9$~K.}}
\end{figure}

The radiative neutron capture cross section is estimated within the statistical
model of Hauser-Feshbach making use of the MOST code \cite{go01a}. This version
benefits in particular from an improved description of the nuclear ground state
properties derived from the microscopic Hartree-Fock method \cite{go01b}, as
well as from a reliable nuclear level density prescription based on the
microscopic statistical model \cite{dem01}. The direct capture contribution as well as
the possible overestimate of the statistical predictions for resonance-deficient nuclei
are effects that could have an important impact on the radiative neutron captures by
exotic nuclei
\cite{go98}, but will not be included in the present analysis. We show in
Fig.~\ref{fig06} for the Sn isotopes, the ratio of the radiative neutron capture
cross section (at an arbitrary incident energy of 10~MeV exemplifying a high energy
regime for which the statistical approximation of the Hauser-Feshbach model remains
valid) predicted with the
$E1$-strength taken from the QRPA to the one taken from the Hybrid Lorentzian-type
formula
\cite{go98}. 
The temperature-dependent Hybrid formula corresponds to a generalization of
the energy- and temperature-dependent Lorentzian formula including an improved
description  of the
$E1$-strength function at energies below $S_n$ as derived from \cite{ka83}. The
strength of the hybrid formula differs from the QRPA estimate not only in the location
of the centroid energy, but also in the low-energy tail determined by the temperature
and energy-dependent width. No extra low-lying strength is included in the
phenomenological Hybrid formula, but its temperature dependence increases the E1
strength at low energies and is responsible for its non-zero $E \rightarrow 0$ limit.
The newly-derived $E1$-strength gives an increase of the cross section by a
factor up to 20 close to the neutron drip line. This is due to the shift of
the GDR to lower energies compared with the usually adopted $A^{-1/3}$ rule (see
Sect.~1 and Fig.~\ref{fig05}), as well as to the
appearance of the GDR strength at low energies as explained above. Both effects tend to
increase the $E1$ strength at energies below the GDR, i.e in the energy window of
relevance in the neutron capture process. For less exotic nuclei, the QRPA impact
is relatively small, differences being mainly due to the exact position of the GDR energy
and the resulting low-energy tail, i.e the energy-dependent width. 

The Maxwellian-averaged $(n,\gamma)$ rates at a temperature of
$1.5~10^9$~K characteristic of the r-process nucleosynthesis are also displayed in
Fig.~\ref{fig06}. These rates are sensitive to the neutron capture cross section at
incident energies around 130 keV, and therefore depend on the $E1$-strength in an narrow
range of a few hundred keV around $S_n$. For nuclei with low neutron
separation energies, the resonant contribution around $S_n$ can be negligible,
so that the stellar rate is principally dominated by captures of a few MeV neutrons.
Deviations up to a factor of 5 are found with respect to the Hybrid
formula (Fig.~\ref{fig06}). The GDR strength predicted at low energies (see
Fig.~\ref{fig05}) is mainly responsible for the QRPA increase of the stellar reaction
rate. In particular, below the neutron shell closure at $N=82$, the neutron separation
energy is large enough (5--7~MeV), so that the GDR tail resulting from the low-energy
QRPA strength enhances the reaction rate by a factor of 2. Above $N=82$, the neutron
separation energy drops to low values ($S_n \lsimeq 3$~MeV) and the reaction rates
increases proportionnally to the $E1$ strength shifted to lower energies for the most
exotic nuclei. For example in the
$^{144}{\rm Sn}(n,\gamma)^{145}{\rm Sn}$ case, the stellar rate and the cross section at
10~MeV are reduced by a factor of 2 and 3, respectively, if the $E1$ strength below
7~MeV is neglected (Fig.~\ref{fig05}). These results show qualitatively the important
effect the extreme neutron excess can have on the neutron capture cross section.
\begin{figure}
\centerline{\epsfig{figure=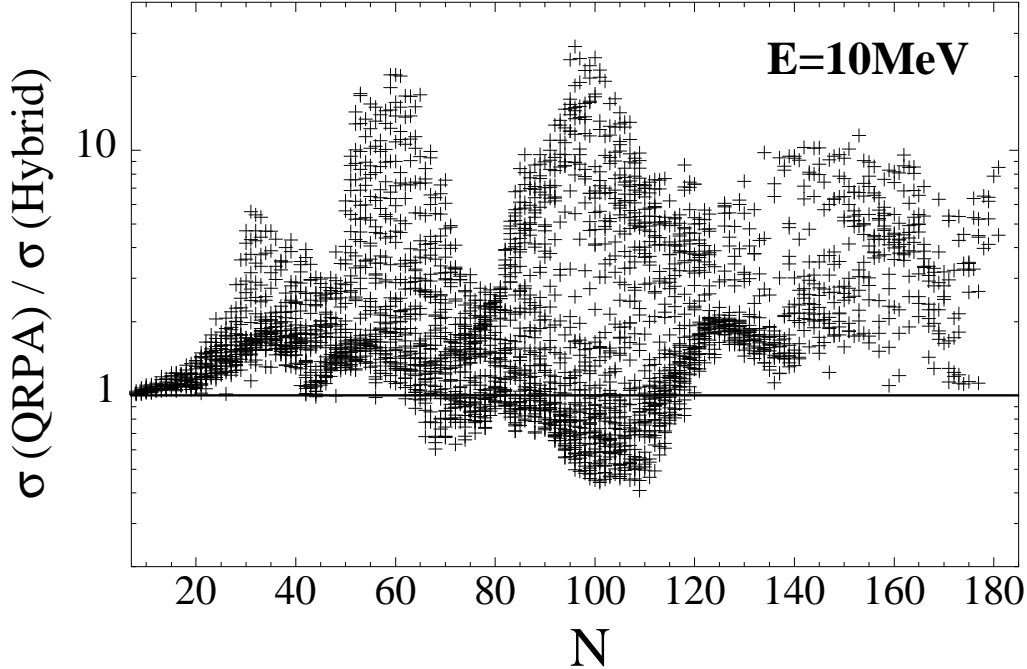,height=9.0cm,width=14cm}}
\caption{\label{fig07}{Ratio of the $(n,\gamma)$ cross section (at $E=10$~MeV) predicted
with the QRPA $E1$-strength to the one derived using the Hybrid formula
\cite{go98}. The figures include about 4000 nuclei with  $8\le Z
\le 84$ lying between the neutron and the proton drip lines and are plotted as a
function of the neutron number. }}
\end{figure}
\begin{figure}
\centerline{\epsfig{figure=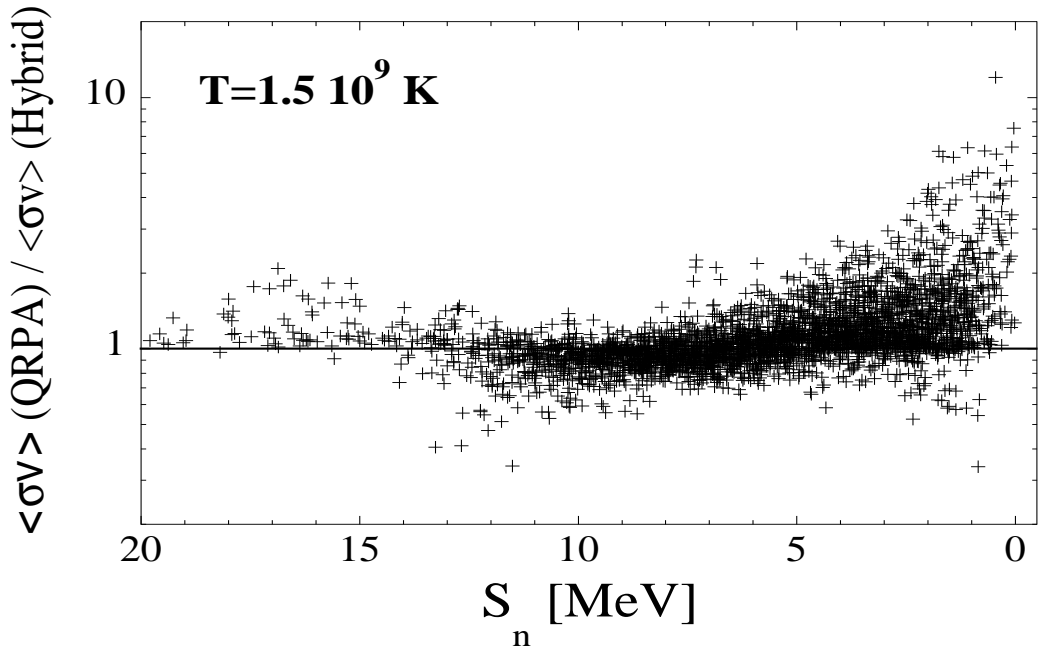,height=9.0cm,width=14cm}}
\caption{\label{fig08}{Same as Fig.~\ref{fig07} for the Maxwellian-averaged neutron
capture rate (at a temperature of $1.5~10^9$~K) as a function of the neutron separation
energy $S_n$.}}
\end{figure}

Similarly, the QRPA calculations have been used to estimate the neutron capture
cross sections and Maxwellian-averaged rates for the 4000 nuclei with $8\le Z
\le 84$ lying between the neutron and the proton drip lines. They are compared in 
Figs.~\ref{fig07}-\ref{fig08} to the predictions based on the Hybrid formula. Increases
up to a factor of 30 are obtained in the cross section at 10~MeV and a factor of 8 in
the rate at $1.5~10^9$~K, in both cases for the most exotic neutron-rich nuclei.
A small decrease of the neutron capture cross sections and rates can also be observed,
especially  in the deformed neutron-deficient region ($S_n\simeq 10$~MeV,  $Z\simeq 70$
and $N\simeq105$) where the neutron separation energy is large and the QRPA predicts, in
the respective energy range of relevance, an $E1$-strength smaller than does the Hybrid
formula. In  this case, the neutron capture cross section with the Hybrid formula can
be up to twice larger than the one obtained with the QRPA strength. This effect is
closely related to the folding procedure adopted here, leading to a zero
limit of the strength function at zero energy in contrast to the temperature-dependence
of the Hybrid prescription leading to a non-zero limit \cite{go98}. Qualitatively,
similar results are obtained with the MSk7 predictions of the $E1$-strength.
Quantitatively, the increase of the neutron capture cross section by nuclei close to the
neutron drip line is slightly smaller with MSk7 than with SLy4. These qualitative
results points out the necessity to perform more predictive calculations, such as
continuum-QRPA, to have a reliable description of nuclei along the drip-line. The impact
of the newly-derived rates on the r-process nucleosynthesis will be studied in a
forthcoming paper.

\section{Conclusions}
The total photon transmission coefficient from a compound nucleus excited
state is one of the key ingredients for statistical cross section
evaluation. The cross section for the radiative capture of low-energy
neutrons strongly depends on the low-energy tail of the giant dipole resonance.  Recent
experimental and theoretical work predict a particular enhancement of the $E1$-strength
at low energies, in particular for neutron-rich nuclei. This low-energy
component of the $E1$-strength can only be derived from microscopic
calculations. Microscopic descriptions of the $E1$-strength function have become
available, mainly in the QRPA approach. None of the previous QRPA calculations, however,
provides a complete set of $E1$-strength functions to be used for practical applications.

For these reasons, we present here the first attempt to construct a complete set of
$E1$-strength function derived from QRPA calculations based on an effective
nucleon-nucleon interaction of Skyrme-type. The QRPA predictions using the SLy4
Skyrme force are in close agreement with photoabsorption data. A folding
procedure is applied to the QRPA strength to take the damping of the collective motion
into account. A prescription for the
folding procedure and for the deformation effects is proposed and shown to lead to
$E1$-strength functions in fair agreement with the available experimental data, in
particular at low energies.
 
The folded QRPA strength is used for the evaluation of the neutron capture
cross section within the framework of the statistical model of
Hauser-Feshbach. The cross sections at energies around 10~MeV are found to
be enhanced by a factor up to 30 when approaching the neutron drip line. The
low-energy contribution to the $E1$ strength predicted by the QRPA
calculations is held responsible for such an increase. No empirical
prescription of the $E1$-strength as traditionally used in reaction
calculation can predict this increase of the radiative capture. This
emphasizes the crucial role of nuclei along the drip-lines, and more
predictive calculations in this area are called for. The neutron capture
rate at temperatures of relevance to nucleosynthesis applications is however
less sensitive to the GDR strength calculated here. Increase up to a factor
of 8 are obtained. The QRPA strength is tabulated and made available to the
scientific community via the nuclear astrophysics library at {\it
http://www-astro.ulb.ac.be}.

Further improvements might be necessary for a better description of the
low-energy tail of the GDR. In particular, the particle-vibration coupling is also known
to affect the low-energy strength and could contribute to an extra increase of the
radiative neutron capture rate by exotic nuclei. A more fundamental determination of the
GDR width, i.e of the damping of the collective motion, as well as a more consistent
treatment of deformation effects also remain to be studied in the future. A
proper treatment of the pairing and the coupling to the continuum for nuclei
along the drip-line is also crucial to have a quantitative idea of the
increase of the neutron capture cross-sections. 
\\

\smallskip
\noindent{\bf Acknowledgments} S.G. is FNRS Research Associate. This work has been
performed within the  scientific collaboration (Tournesol) between the
Wallonie--Bruxelles Community and France. \\

\end{document}